\documentclass[reprint,doublecolumn,secnumarabic,amssymb,preprintnumbers, amsmath, aps,nofootinbib, superscriptaddress]{revtex4-1}
\pdfoutput=1

\usepackage{amsmath}
\usepackage{amsfonts}
\usepackage{amsthm}
\usepackage{amssymb}
\usepackage{latexsym, array,multirow,  verbatim, enumerate}
\usepackage{slashed}
\usepackage{cancel}
\usepackage{dcolumn}
\usepackage{bm}
\usepackage{graphicx, caption, subcaption}  
\usepackage{url}
\usepackage[normalem]{ulem}
\usepackage{enumitem}
\usepackage{bbding}
\usepackage{tikz}
\usepackage{pifont}
\usepackage{empheq}
\usepackage[utf8]{inputenc}
\usepackage{bm}

\usepackage[colorlinks]{hyperref}

\usepackage{ulem}

\begin{document}
\title{Neutrino  charge constraints from scattering to the weak gravity conjecture to neutron stars}

 \author{Arindam Das}
 \email{arindam.das@het.phys.sci.osaka-u.ac.jp}
\affiliation{Osaka University, Toyonaka, Osaka 560-0043, Japan}
\author{Diptimoy Ghosh}
\email{diptimoy.ghosh@iiserpune.ac.in}
\affiliation{Department of Physics, Indian Institute of Science Education and Research, Pune 411008, India}
\author{Carlo Giunti}
\email{carlo.giunti@to.infn.it}
\affiliation{INFN, Torino Section, Via P. Giuria 1, I-10125 Torino, Italy}
\author{Arun Thalapillil}
\email{thalapillil@iiserpune.ac.in}
\affiliation{Department of Physics, Indian Institute of Science Education and Research, Pune 411008, India}
\preprint{OU-HET-1059}
\date{\today}
\begin{abstract}
In various extensions of the Standard Model of particle physics, and intriguingly even in the three-generation Standard Model without neutrino masses, neutrinos are allowed to have very tiny electric charges. After a review of the theoretical scenarios that allow the emergence of such charges, we discuss the existing observational limits and we derive new stringent direct upper bounds for the charges of the muon and tau neutrinos. We also point out a flavor-universal lower bound on neutrino charges which is obtained from the weak gravity conjecture, that is based on the hypothesis that gravity is the weakest force. We finally present a new flavor-universal upper bound on neutrino charges
based on astrophysical observations of Magnetars.
\end{abstract}

\maketitle

\section{Introduction}
\label{sec:intro}

The Standard Model of particle physics (SM) has been a remarkably successful theory in both predictive power and breadth of applicability. Nevertheless, there is a veritable cornucopia of evidence to suggest that the SM is incomplete and must be extended. In the context of our study, neutrinos may provide a window to some of these aspects. In fact, the neutrino sector may be an important participant in potential solutions to baryon asymmetry in the universe, CP violation in nature and the fermion mass hierarchy. Moreover, some extensions of the SM may even embody neutrinos with novel electromagnetic properties through quantum loop effects, making them complementary probes for beyond-SM physics (see for instance\,\cite{Giunti:2014ixa,Giunti:2020dqw} and related references). 

In the SM, as it is usually understood, neutrinos are strictly massless. However, the observation of neutrino flavor oscillations\,\cite{Abe:2010hy,Aharmim:2008kc,Tanabashi:2018oca} suggest that at least two of the neutrinos are massive. This can be achieved by the introduction of right-handed neutrinos, and corresponding Dirac and Majorana mass terms, or through the see-saw framework \cite{Minkowski:1977sc,Mohapatra:1979ia,Schechter:1980gr,Yanagida:1979as,GellMann:1980vs,Glashow:1979nm}. 

The issue of a non-zero neutrino electric charge is much more complicated and interesting (see for instance \,\cite{Babu:1989ex, Babu:1989tq, Foot:1990uf,Foot:1992ui,Giunti:2014ixa} and related references). Indeed, even in the SM (with three generations) their electric charges are not fully determined just by the mathematical consistency of the SM as a quantum field theory, viz. the lack of gauge anomalies and possibly also the mixed gauge-gravitational anomaly \cite{Delbourgo:1972xb,Eguchi:1976db,AlvarezGaume:1983ig,Geng:1988pr,Byakti:2017igi} (see section~\ref{sec:models} for more details). Thus, whether neutrinos are electrically charged is entirely an experimental question. As we review in section~\ref{sec:models}, interestingly, the existence of neutrino charges is related to the nature of neutrino masses\,\cite{Babu:1989ex,Babu:1989tq} (i.e., whether one has Dirac or Majorana mass terms in the Lagrangian density), and whether or not electric charges are quantized in nature. In almost all cases, the dequantization of electric charges is closely related to the emergence of additional non-anomalous abelian symmetries.

There is no strict upper bound on a possible neutrino electric charge, as far as purely theoretical considerations go. Any upper bound is instead motivated by experimental and observational considerations. The non-neutrality limits on matter put a strong bound on the electric charge of the electron-type neutrino. Apart from that, there are also several limits coming from low energy reactor neutrinos, beam dump experiments, galactic-extragalactic neutrino sources, stellar cooling, neutrino star turning mechanism and so on. These existing constraints are briefly reviewed in sections~\ref{sec:labconstraints} and~\ref{sec:astroconstraints}. In section~\ref{sec:newconstraints} we derive new direct bounds on the charges of $\nu_\mu$ and $\nu_\tau$. If neutrinos have a non-zero electric charge, one can derive a possible theoretical lower bound on their charge-to-mass ratio from the (strong form) weak gravity conjecture \cite{ArkaniHamed:2006dz}. We discuss this in section~\ref{sec:wgc}. In section~\ref{sec:astrospp}
we demonstrate a new flavour-universal upper bound on neutrino charges that is deduced from the observation of neutron stars with very large magnetic fields (Magnetars), where charged neutrinos can be non-perturbatively pair-produced, thereby contributing to a depletion of the energy reservoir.

\section{Neutrino charge in the SM and beyond}
\label{sec:models}

Let us briefly review and discuss scenarios where neutrino electric charges may arise naturally. We use the convention where all the SM fermions are left-chiral. They are denoted by $Q(3,2), U^c(\bar{3},1), D^c(\bar{3},1), L(1,2),$ and $E^c(1,1)$ where the numbers in parentheses
denote the $SU(3)_c$ and $SU(2)_L$ charges. The Higgs field is denoted by $H(1,2)$.

The hypercharge quantum numbers of the SM fields must obey the following consistency conditions due to anomaly cancellation requirements:
\begin{itemize}
\item
from the $U(1)_Y$-$[SU(3)_c]^2$ anomaly
\begin{equation}
\sum_{i=1}^{3} \left[ 2Y_Q^{(i)} + Y_{U^c}^{(i)} + Y_{D^c}^{(i)} \right] = 0
;
\label{U1Su3Su3}
\end{equation}
\item
from the $U(1)_Y$-$[SU(2)_L]^2$ anomaly
\begin{equation}
\sum_{i=1}^{3} \left[ 3Y_Q^{(i)} + Y_{L}^{(i)} \right] = 0
;
\label{U1SU2SU2}
\end{equation}
\item
from the $[U(1)_Y]^3$ anomaly
\begin{align}
\sum_{i=1}^{3}
\left[ 2 \left(Y_L^{(i)}\right)^3 + \left(Y_{E^c}^{(i)}\right)^3 + 6 \left(Y_Q^{(i)}\right)^3 \right.
&
\nonumber
\\
+
\left. 3 \left(Y_{U^c}^{(i)}\right)^3 + 3 \left(Y_{D^c}^{(i)}\right)^3 \right]
&
\null = 0
;
\label{U1U1U1}
\end{align}
\item
from the $U(1)_Y$-$[\text{graviton}]^2$ anomaly
\begin{equation}
\sum_{i=1}^{3}
\left[ 2 Y_L^{(i)} + Y_{E^c}^{(i)} + 6 Y_Q^{(i)}
+ 3 Y_{U^c}^{(i)} + 3 Y_{D^c}^{(i)} \right]
= 0
.
\label{U1GG}
\end{equation}
\end{itemize}
The hypercharge quantum numbers must also obey the following constraints from the gauge invariance
of the Yukawa terms in the Lagrangian density:
\begin{itemize}
\item
from the charged-lepton Yukawa terms
\begin{equation}
Y_L^{(i)} + Y_{E^c}^{(i)} = Y_H \quad \forall i
;
\label{Yukl}
\end{equation}
\item
from the up-type quark Yukawa terms
\begin{equation}
Y_Q^{(i)} + Y_{U^c}^{(j)} = - Y_H \quad \forall \{i, j\}
;
\label{Yuku}
\end{equation}
\item
from the down-type quark Yukawa terms
\begin{equation}
Y_Q^{(i)} + Y_{D^c}^{(j)} = Y_H \quad \forall \{i, j\}
.
\label{Yukd}
\end{equation}
\end{itemize}
In the above equations
the superscript indices $\{i,j\}$ indicate the generation.
$Y_H$ is the hypercharge of the Higgs doublet.
Note also that the charged-lepton Yukawa has been taken to be diagonal
considering the SM where neutrinos are massless and there is no mixing in the lepton sector.
On the other hand, the mixing in the quark sector imposes the relations (\ref{Yuku}) and (\ref{Yukd})
between the quark hypercharges of different generations,
that imply straightforwardly that the quark hypercharges are generation-independent:
$Y_Q^{(i)} = Y_Q$,
$Y_{U^c}^{(i)} = Y_{U^c} = - Y_H - Y_Q$, and
$Y_{D^c}^{(i)} = Y_{D^c} = Y_H - Y_Q$.
Moreover, inserting these equalities in Eq.~(\ref{U1Su3Su3})
one can see that the $U(1)_Y$-$[SU(3)_c]^2$ anomaly cancels automatically
and does not imply any further constraint.
Also the quark contributions to the mixed gauge-gravitational anomaly
cancel and Eq.~(\ref{U1GG}) yields the simpler relation between lepton hypercharges
$
\sum_{i=1}^{3}
\left[ 2 Y_L^{(i)} + Y_{E^c}^{(i)} \right]
= 0
$.
Then, using also the charged-lepton Yukawa constraint (\ref{Yukl})
and the $U(1)_Y$-$[SU(2)_L]^2$ anomaly (\ref{U1SU2SU2}),
we obtain the relations
\begin{align}
&
Y_Q = \dfrac{1}{3} \, Y_H
,
\label{YQ}
\\
&
Y_{U^c} = - \dfrac{4}{3} \, Y_H
,
\label{YU}
\\
&
Y_{D^c} = \dfrac{2}{3} \, Y_H
,
\label{YD}
\\
&
Y_L^{(i)} = Y_H - Y_{E^c}^{(i)} \quad \forall i
,
\label{YL}
\\
&
\sum_{i=1}^{3} Y_{E^c}^{(i)} = 6 \, Y_H
.
\label{YE}
\end{align}

In the case of the one-generation SM, one gets $Y_{E^c} = 2Y_H$ and consequently, $Y_L = - Y_H$.
In this case, all the hypercharges are given in terms of $Y_H$, which result in quantized hypercharges (i.e. the ratios of all charges are rational numbers).
Moreover, the $[U(1)_Y]^3$ anomaly constraint (\ref{U1U1U1})
is automatically satisfied for any value of $Y_H$
and does not imply any further constraint.
Setting $Y_H = +1$, one can recover the conventional values of hypercharges usually given in standard textbooks:
\begin{equation}
Y_Q^{\text{SM}} = \dfrac{1}{3}
, \,
Y_{U^c}^{\text{SM}} = - \dfrac{4}{3}
, \,
Y_{D^c}^{\text{SM}} = \dfrac{2}{3}
, \,
Y_L^{\text{SM}} = -1
, \,
Y_{E^c}^{\text{SM}} = 2
.
\label{YSM}
\end{equation}
Note that since $Y_H$ is an overall scaling of all hypercharges, it can always be set to unity by a suitable choice of the hypercharge gauge coupling. However, we will keep it as a free parameter to be as general as possible, as far as expressions are concerned. In this case, the Gell-Mann--Nishijima formula gets modified to
\begin{equation}
Q=Y_H I_3+\frac{Y}{2} \; .
\end{equation}
The electric charge of the neutrino can now be computed as 
\begin{equation}
Q_{\nu}^{(i)}= Y_H  - \frac{Y_{E^c}^{(i)}}{2}  \; .
\end{equation}
Thus, we deduce that in the one-generation SM, when $Y_{E^c}=2Y_H$ in Eqs.~(\ref{YL}) and (\ref{YE}),
the neutrinos are exactly neutral:
\begin{equation}
Q_{\nu} = 0 ~;~~(\text{One-generation SM})\; .
\end{equation}

On the other hand, in the three-generation SM, with massless neutrinos, the electric charge is not quantized
and not all the neutrinos have to be neutral~\cite{Foot:1990uf,Foot:1992ui}.
Indeed,
Eq.~(\ref{YE}) leaves freedom for the individual value of each of the three hypercharges $Y_{E^c}^{(i)}$,
that implies a corresponding freedom for the individual value of each of the three hypercharges $Y_L^{(i)}$
through Eq.~(\ref{YL}).
However, in this case we must take into account that the hypercharges are further constrained by
the $[U(1)_Y]^3$ anomaly constraint (\ref{U1U1U1}),
that gives
\begin{equation}
\sum_{i=1}^{3} \left[ \left(Y_{E^c}^{(i)}\right)^3 - 6 Y_H \left(Y_{E^c}^{(i)}\right)^2 \right]
+ 48 Y_H^3 = 0
.
\label{Y3}
\end{equation}
In order to understand what is the difference with respect to the one-generation SM,
let us write $Y_{E^c}^{(i)}$ as
\begin{equation}
Y_{E^c}^{(i)} = 2 Y_H \left( 1 + \delta^{(i)} \right)
.
\label{YEdev}
\end{equation}
Then,
Eqs.~(\ref{YE}) and (\ref{Y3}) yield, respectively, the constraints
\begin{align}
&
\sum_{i=1}^{3} \delta^{(i)} = 0
,
\label{d1}
\\
&
\sum_{i=1}^{3} \left(\delta^{(i)}\right)^3 = 0
.
\label{d2}
\end{align}
Therefore, only two of the three $\delta^{(i)}$'s can be different from zero and their values must be opposite:
\begin{align}
&
\delta^{(1)} = - \delta^{(2)}
\quad \text{and} \quad
\delta^{(3)} = 0
,
\quad \text{or}
\label{c12}
\\
&
\delta^{(2)} = - \delta^{(3)}
\quad \text{and} \quad
\delta^{(1)} = 0
,
\quad \text{or}
\label{c23}
\\
&
\delta^{(3)} = - \delta^{(1)}
\quad \text{and} \quad
\delta^{(2)} = 0
.
\label{c31}
\end{align}
As discussed in Refs.~\cite{Foot:1990uf,Foot:1992ui},
this dequantization of the electric charge in the three generation SM is related to the existence of the $U(1)$ symmetries corresponding to the
three differences of the generation lepton numbers:
$(L_e - L_\mu)$, $(L_\mu - L_\tau)$ and $(L_e - L_\tau)$.
Only one of these three $U(1)$ symmetries can be non-anomalous
and the corresponding lepton number difference
can be added to the hypercharge with an arbitrary coefficient,
leading to the generation of the contributions $\delta^{(i)}$ and
the dequantization of the electric charge.
This is a particular example of the general mechanism of charge dequantization
induced by anomaly-free $U(1)$ symmetries~\cite{Holdom:1985ag}.

The electric charges of the neutrinos in the three-generation SM are given by
\begin{equation}
Q_{\nu}^{(i)} = - Y_H \delta^{(i)} ~;~~(\text{Three-generation SM})
.
\label{QnuSM3}
\end{equation}
Only two of the three neutrino electric charges can be non-zero and they have to be opposite, in such a way that
the sum of the neutrino charges vanishes:
\begin{equation}
\sum_{i=1}^{3} Q_{\nu}^{(i)} = 0
.
\label{Qnu3sum}
\end{equation}
For the other charges, in this case, we get
\begin{align}
&
Q_{e}^{(i)} = - Q_{E^c}^{(i)} = - Y_H \left( 1 + \delta^{(i)} \right)
,
\label{QeSM3}
\\
&
Q_{u} = - Q_{U^c} = \frac{2}{3} \, Y_H
,
\label{QuSM3}
\\
&
Q_{d} = - Q_{D^c} = - \frac{1}{3} \, Y_H
.
\label{QdSM3}
\end{align}
Therefore,
only the lepton charges are dequantized and they satisfy the relations
$Q_{e}^{(i)}=Q_\nu^{(i)}-Y_H$.

Let us now consider scenarios where the neutrinos have non-zero masses. The observation of neutrino oscillations\,\cite{Abe:2010hy,Aharmim:2008kc,Tanabashi:2018oca} and the subsequent extraction of the neutrino mass-squared differences\,\cite{Tanabashi:2018oca} imply that at least two of the neutrino mass eigenstates must have non-vanishing masses. This mass may arise either from Dirac mass terms or Majorana mass terms added to the SM Lagrangian.

If one introduces Majorana mass terms for the left-chiral neutrinos through the Weinberg operator, the following additional consistency conditions must be satisfied:
\begin{align}
Y_L^{(i)}+Y_L^{(j)} + 2Y_H = 4Y_H - Y_{E^c}^{(i)} - Y_{E^c}^{(j)} = 0 \quad \forall \{i, j\}
\end{align}
This gives $Y_{E^c}^{(i)} = 2Y_H \, \forall i$ and, consequently, all the neutrinos must be neutral\,\cite{Babu:1989ex,Foot:1990uf}:
\begin{equation}
 Q_{\nu}^{(i)} = 0  ~\forall i~;~~(\text{SM with $\nu$ Majorana mass term})  \; .
\end{equation}
Thus, in this case the charges are again quantized.

Let us now consider the alternative scenario where neutrinos get their masses from a Dirac mass term in the Lagrangian.
This can be achieved with the addition of three copies of a singlet (under $SU(3)_c$ and $SU(2)_L$) neutrino $N^c(1,1)$ and the corresponding Dirac mass terms.
These additions modify the $[U(1)_Y]^3$ and mixed gauge-gravitational anomaly constraints
and
induce a new constraint equation corresponding to the Dirac neutrino Yukawa term:
\begin{itemize}
\item
from the $[U(1)_Y]^3$ anomaly
\begin{align}
\sum_{i=1}^{3}
\left[ 2 \left(Y_L^{(i)}\right)^3 + \left(Y_{E^c}^{(i)}\right)^3 + 6 \left(Y_Q^{(i)}\right)^3 \right.
&
\nonumber
\\
+
\left. 3 \left(Y_{U^c}^{(i)}\right)^3 + 3 \left(Y_{D^c}^{(i)}\right)^3 + \left(Y_{N^c}^{(i)}\right)^3 \right]
&
\null = 0
;
\label{U1U1U1Dirac}
\end{align}
\item
from the $U(1)_Y$-$[\text{graviton}]^2$ anomaly
\begin{equation}
\sum_{i=1}^{3}
\left[ 2 Y_L^{(i)} + Y_{E^c}^{(i)} + 6 Y_Q^{(i)}
+ 3 Y_{U^c}^{(i)} + 3 Y_{D^c}^{(i)} + Y_{N^c}^{(i)} \right]
= 0
;
\label{DU1GG}
\end{equation}
\item
from the Dirac neutrino Yukawa terms
\begin{equation}
Y_L^{(i)} + Y_{N^c}^{(j)} = - Y_H \quad \forall \{i, j\}
.
\label{DYukl}
\end{equation}
\end{itemize}
In principle,
also the charged-lepton Yukawa constraint (\ref{Yukl})
should be modified allowing all possible different indices,
but we do not need to do it because Eq.~(\ref{DYukl}) implies straightforwardly
that all $Y_L^{(i)}$'s are equal.
Let us also note that in this case the mixed gauge-gravitational anomaly constraint (\ref{DU1GG}) is redundant and not necessary,
because it
is automatically satisfied for any value of $Y_H$ using the Yukawa relations.
The new solution of all the constraints is
\begin{align}
&
Y_{N^c}^{(1)}  = Y_{N^c}^{(2)} = Y_{N^c}^{(3)} = Y_{N^c}
,
\label{DYN}
\\
&
Y_Q^{(i)}  = \frac{Y_H}{3} + \frac{Y_{N^c}}{3} \quad \forall i
,
\label{DYQ}
\\
&
Y_{U^c}^{(i)} =-\frac{4Y_H}{3} - \frac{Y_{N^c}}{3} \quad \forall i
,
\label{DYU}
\\
&
Y_{D^c}^{(i)}  = \frac{2Y_H}{3} - \frac{Y_{N^c}}{3} \quad \forall i
,
\label{DYD}
\\
&
Y_L^{(i)} = -Y_H - Y_{N^c} \quad \forall i
,
\label{DYL}
\\
&
Y_{E^c}^{(i)} = 2 Y_H + Y_{N^c} \quad \forall i
.
\label{DYE}
\end{align}
Hence,
in this case all the hypercharges and the corresponding quark and lepton electric
charges are generation-independent.
They are also not completely determined, since they depend on the arbitrary hypercharge assignment for the right-handed neutrino $Y_{N^c}$.
In this case, the neutrino charge is given by
\begin{equation}
Q_{\nu}^{(i)} = - \frac{Y_{N^c} }{2} ~\forall i~;~~(\text{SM with $\nu$ Dirac mass term})  \; .
\end{equation}
The other electric charges are given by
\begin{align}
&
Q_{e} = - Q_{E^c}= - Y_H - \frac{Y_{N^c}}{2}
,
\label{DQe}
\\
&
Q_{u} = -Q_{U^c} = \frac{2}{3} \, Y_H  + \frac{Y_{N^c}}{6}
,
\label{DQu}
\\
&
Q_{d} = -Q_{D^c} = - \frac{1}{3} \, Y_H  + \frac{Y_{N^c}}{6}
.
\label{DQd}
\end{align}
We see that in this case the charged lepton, up-type quark and down-type quark electric charges are related to $Q_\nu$ by
\begin{align}
&
Q_{e} = - Y_H + Q_\nu
,
\label{DQenu}
\\
&
Q_{u} = \frac{2}{3} \, Y_H  - \frac{Q_\nu}{3}
,
\label{DQunu}
\\
&
Q_{d} = - \frac{1}{3} \, Y_H  - \frac{Q_\nu}{3}
.
\label{DQdnu}
\end{align}
Thus, in the presence of right-handed neutrinos and Dirac mass terms for the neutrinos, electric charge is 
dequantized and neutrinos can be electrically charged\,\cite{Babu:1989ex}. The charge dequantization in this case is related to the
existence of the non-anomalous symmetry $(B-L)$\,\cite{Babu:1989ex,Foot:1990uf}. Note that in this case, neutrinos can be charged even when there is only one generation of fermions. 

\section{Laboratory constraints on neutrino charges}
\label{sec:labconstraints}

A variety of experimental and observational considerations constrain possible neutrino charges. These bounds come from both terrestrial as well as astrophysical observations. In this Section we briefly review the existing laboratory constraints on the neutrino charges.

The strongest experimental constraint on first-generation neutrinos is obtained from beta decay $n\to p+e^-+\overline{\nu}_e$, in combination with limits on the non-neutrality of matter. The neutrality of the matter is usually quantified in terms of $Q_{\text{matter}} = \frac{1}{A} [Z(Q_p+ Q_e)+ (A-Z) Q_n]$. Here, $Z$ is the atomic number, $N$ is the neutron number and $A$ is the atomic mass number of the element.
$Q_p$, $Q_e$ and $Q_n$ are the electric charges of the proton, electron and neutron respectively. Conservation of the electric charge in beta decay requires $Q_{\nu_e} = \frac{A (Q_n-Q_{\text{matter}})}{Z}$. The non-neutrality test of matter\,\cite{Bressi:2011yfa} using Sulphur Hexafluoride (SF$_6$) sets a strong bound $Q_{\text{matter}}= (-0.1\times1.1)\times 10^{-21} e$. The independent measurement of the charge of a free neutron sets a limit of $Q_n=(-0.4 \pm 1.1)\times 10^{-21}e$\,\cite{Raffelt:1999gv,Giunti:2014ixa}. Both of these in combination then puts the strong constraint~\cite{Raffelt:1999gv,Giunti:2014ixa}
\begin{equation}
Q_{\nu_e} = \left( -0.6 \pm 3.2 \right) \times 10^{-21} \, e
.
\label{neutrality}
\end{equation}
This bound should also be applicable to the other generation of neutrinos when neutrino flavor oscillations are taken into account.

The electric charge of neutrinos was also probed directly in scattering experiments.
From the TEXONO experiment, low-energy reactor antineutrino scattering with electrons provide the 90\% CL upper bound~\cite{Chen:2014dsa}
\begin{equation}
|Q_{\nu_e}| < 2.1 \times 10^{-12} \, e
.
\label{TEXONO}
\end{equation}
An improved upper bound on $Q_{\nu_e}$
has been obtained~\cite{Studenikin:2013my}
using the most stringent bound on the electron neutrino magnetic of the
GEMMA collaboration:
$ \mu_{\nu_e} < 2.9 \times 10^{-11} \, \mu_{\mathrm{B}} $
at 90\% CL~ \cite{Beda:2009kx},
where $\mu_{\mathrm{B}}$ is the Bohr magneton.
A comparison of the cross sections of neutrino--electron
scattering due to a neutrino electric charge and a neutrino magnetic moment
lead to the relation
\begin{equation}
|Q_{\nu}|
\lesssim
\sqrt{\dfrac{T_{e}^{\text{th}}}{2 m_{e}}}
\left( \dfrac{ \mu_{\nu}^{\text{ub}} }{ \mu_{\mathrm{B}} } \right)
\,
e
,
\label{relation}
\end{equation}
where $T_{e}^{\text{th}}$ is the electron kinetic energy threshold
and
$\mu_{\nu}^{\text{ub}}$ is the upper bound for the magnetic moment.
Using $T_{e}^{\text{th}}=2.8\,\text{keV}$,
the 90\% CL GEMMA limit on $\mu_{\nu_e}$ implies the stringent limit
\begin{equation}
|Q_{\nu_e}| \lesssim 1.5 \times 10^{-12} \, e
.
\label{Studenikin}
\end{equation}
An updated and similar bound was also obtained recently, by analyzing  combined data from various elastic neutrino-electron scattering measurements\cite{Parada:2019gvy}, utilizing reactor neutrinos. In the upgraded phase of GEMMA it is expected that one may be able to improve these upper bounds by an order of magnitude\,\cite{Studenikin:2013my}.

The recent first measurements of coherent neutrino-nucleus elastic scattering
(CE$\nu$NS)
in the COHERENT experiment~\cite{Akimov:2017ade,Akimov:2020pdx}
led to the following new constraints on the neutrino electric
charges~\cite{Cadeddu:2019eta,Cadeddu:2020lky}:
\begin{align}
&
Q_{\nu_e} = \left( 10 \pm 14 \right) \times 10^{-8} \, e
,
\label{CENNS-e}
\\
&
Q_{\nu_\mu} = \left( - 1.5 \pm 5.5 \right) \times 10^{-8} \, e
.
\label{CENNS-mu}
\end{align}
The bound on
$Q_{\nu_e}$ is not competitive with the reactor bounds
(\ref{TEXONO}) and (\ref{Studenikin}),
but the bound on $Q_{\nu_\mu}$ was
the only existing one obtained from scattering experiments (see the new bound in the next Section).
The analyses of the COHERENT data in Ref.~\cite{Cadeddu:2019eta,Cadeddu:2020lky}
constrained also the transition electric charges,
that contribute to the scattering~\cite{Kouzakov:2017hbc}:
at $3\sigma$
\begin{align}
&
|Q_{\nu_{e\mu}}| < 20 \times 10^{-8} \, e
,
\label{CENNS-e-mu}
\\
&
|Q_{\nu_{e\tau}}| < 34 \times 10^{-8} \, e
,
\label{CENNS-e-tau}
\\
&
|Q_{\nu_{\mu\tau}}| < 25 \times 10^{-8} \, e
.
\label{CENNS-mu-tau}
\end{align}

The SLAC electron beam dump experiment provided the following upper limit on the third-generation neutrino $(\nu_\tau)$ charge~\cite{Davidson:1991si}:
\begin{equation}
|Q_{\nu_\tau}| \lesssim 3 \times 10^{-4} \, e
.
\label{SLAC}
\end{equation}
Beam dump experiments utilizing bubble chambers, such as BEBC\,\cite{CooperSarkar:1991xz}, have also constrained the charge of $\nu_\tau$ from the elastic scattering $\nu_\tau e^- \to \nu_\tau e^-$. Comparison of the theoretical expectation, for the scattering cross section (proportional to $Q_{\nu_{\tau}}^2$), with the experimental observation, provided the upper bound~\cite{Babu:1993yh}
\begin{equation}
|Q_{\nu_\tau}| \lesssim 4 \times 10^{-4} \, e
.
\label{BEBC}
\end{equation}

\section{New upper bounds for $|Q_{\nu_\mu}|$ and $|Q_{\nu_\tau}|$}
\label{sec:newconstraints}

We have seen in the previous Section that using the
relation (\ref{relation}) one can convert an upper bound for a neutrino magnetic moment
into an upper bound for the electric charge.
In this section we achieve new direct bounds on
$|Q_{\nu_\mu}|$ and $|Q_{\nu_\tau}|$
by applying this method to the most stringent laboratory constraints on the
magnetic moments of the muon and tau neutrino.

Let us first consider the most stringent bound
$ \mu_{\nu_\mu} < 6.8 \times 10^{-10} \, \mu_{\mathrm{B}} $
(90\% CL)
on the muon neutrino magnetic moment
obtained in the LSND experiment~\cite{Auerbach:2001wg}
with neutrino-electron scattering.
Considering the LSND electron energy threshold
$ T_{e}^{\text{th}} = 18 \, \text{MeV} $,
using the relation (\ref{relation})
we obtain the following new upper bound on the electric charge of the muon neutrino:
\begin{equation}
|Q_{\nu_\mu}| \lesssim 2.9 \times 10^{-9} \, e
.
\label{LSND}
\end{equation}
This limit is stronger than the previously most stringent direct limit in Eq.~(\ref{CENNS-mu}).

Let us now consider
the most stringent bound
$ \mu_{\nu_\tau} < 3.9 \times 10^{-7} \, \mu_{\mathrm{B}} $
(90\% CL)
on the tau neutrino magnetic moment
obtained in the DONUT experiment~\cite{Schwienhorst:2001sj}
with $\nu_\tau$-$e$ scattering.
From the DONUT electron energy threshold
$ T_{e}^{\text{th}} = 0.1 \, \text{GeV} $,
using the relation (\ref{relation})
we obtain
\begin{equation}
|Q_{\nu_\tau}| \lesssim 3.9 \times 10^{-6} \, e
.
\label{DONUT}
\end{equation}
This bound is much stronger than the previously existing direct limits in
Eqs.~(\ref{SLAC}) and (\ref{BEBC}).

\section{Astrophysical constraints on neutrino charges}
\label{sec:astroconstraints}

Astrophysical observations also place various constraints on neutrino charges. While some are relatively model independent, others depend on reasonable assumptions. In this Section we briefly review the main existing astrophysical bounds.

SN1987A supernova neutrino measurements can be used to constrain first-generation neutrino charges\,\cite{Barbiellini:1987zz}. The basic idea is that galactic and extragalactic magnetic fields can cause energy dependence in the arrival times of the charged neutrinos. The SN1987A observations put an upper bound of
\begin{equation}
|Q_{\nu_e}|\lesssim 10^{-15} e- 10^{-17}e \; ,
\end{equation}
depending on the precise value of the mean magnetic field encountered during traversal. 

A very interesting constraint on neutrino charges may be obtained by considering their effects on the rotation of magnetized neutron stars--- the neutrino star turning mechanism ($\nu$ST)\,\cite{Studenikin:2012vi}. The charged neutrinos produced in the stellar interior and traveling out of the rotating, magnetized nuclear matter of the star could potentially slow down its rotation. Hence, charged neutrinos may prevent the generation of a rapidly rotating neutron star, or pulsars may be affected by a frequency shift due to the $\nu$ST mechanism. Considering a magnetic field of $10^{14}$ G and solving the Dirac equation with an ansatz for the magnetized nuclear matter in the neutron star, a strong upper limit of about\,\cite{Studenikin:2012vi}
\begin{equation}
|Q_\nu| \lesssim 10^{-19} e \; ,
\label{eq:nust}
\end{equation}
was obtained. 

If neutrinos are charged, they could also participate in plasmon decays. Considering plasmon decay $\gamma^*\rightarrow \nu \bar{\nu}$ in the sun, and requiring that the energy loss be lower than the solar luminosity, imposes constraints on the neutrino charge. The upper limits obtained from helioseismological studies are about\,\cite{Raffelt:1999gv}
\begin{equation}
|Q_\nu| <  6\times10^{-14} e \; .
\label{eq:shb}
\end{equation}
Such non-standard losses would also delay the ignition of helium in  low-mass red giant cores. From globular-cluster stars, these considerations put a limit of\,\cite{Raffelt:1999gv}
\begin{equation}
|Q_\nu| <  2\times10^{-14} e \; .
\label{eq:scb}
\end{equation}
 
\section{Lower bound on charge-to-mass ratio for electrically charged neutrinos}
\label{sec:wgc}

The weak gravity conjecture (WGC), is an assertion about the relative strength of gravity with respect to gauge forces, in a consistent theory of quantum gravity\,\cite{ArkaniHamed:2006dz}. It crudely pronounces that gravity must be the weakest force in such a theory. More precisely, it states that for a U(1) gauge theory coupled to gravity there must exist at least one 
state with charge $q$ and mass $m$ such that $q > m/\sqrt{2} $ in appropriate units (the units are chosen such that for an extremal black hole with charge $Q$ and mass $M$, $M=\sqrt{2} Q$).
In the case of electromagnetism, in physical units, this can be written as
\begin{eqnarray}
\frac{q}{e}  \geqslant \frac{1}{\sqrt{2}\sqrt{4 \pi \alpha_{em}}} \frac{m}{m_{\text{\tiny{planck}}}} \; ,
\end{eqnarray}
where $m_{\text{\tiny{planck}}}$ is the reduced Planck mass, with $m_{\text{\tiny{planck}}} \approx 2.4 \times 10^{18}$ GeV.

As discussed in\,\cite{ArkaniHamed:2006dz}, the above minimal form of the conjecture, often called the ``electric WGC" does not really impose any interesting constraints on the particle spectrum at low energies. Thus, the so-called strong form of the WGC was proposed which says that the criterion $q > m/\sqrt{2}$ should be satisfied for the lightest charged particle in the spectrum. 

We will ask what the strong form of the WGC may tell us about electrically charged neutrinos. Assume that the neutrino mass eigenstates all have non-vanishing masses. Then, the above considerations gives rise to the following lower bound on the electric charge of the lightest charged neutrino 
\begin{equation}
|Q_\nu| \gtrsim 10^{-28} e\, \left( \frac{m_\nu}{\mathrm{0.1eV}}\right) \, .
\label{eq:strongwgc}
\end{equation}
Hence, in contrast to the existing observational and experimental constraints which furnish an upper bound, theoretical considerations encouraged by the strong form of the WGC indicate a lower bound for the neutrino charge. Note that, there is no formal proof of the
WGC, and in fact, there are considerable debates on whether the strong form of the WGC has to be satisfied by effective field
theories at low energies, see \cite{Palti:2019pca} for a review. Thus, the bound of Eq.~\eqref{eq:strongwgc} should only be taken
as a theoretical guidance, and not as a strict lower bound. 


\section{Constraints from pair production in neutron stars}
\label{sec:astrospp}

Magnetars\,\cite{1992ApJ...392L...9D,1993ApJ...408..194T,Thompson:1995gw} are a class of neutron stars that have extremely large magnetic fields, $10^{14}-10^{15}\,\mathrm{G}$ or higher, and have large mean spin periods, $t_{\text{\tiny{period}}} \sim \mathcal{O}(10)\,\mathrm{sec}$. There is presently overwhelming observational evidence for such objects\,\cite{Olausen:2013bpa}, with characteristic lifetimes $\sim 10^{4}\,\mathrm{yrs}$. 

Neutron stars have a magnetospheric region, with a plasma density, enveloping them. Many models generically predict the existence of acceleration or vacuum gap regions in the neutron star magnetosphere, where the plasma density is very low or vanishing\,\cite{Michel:1982fj, Becker:2009, 2004hpa..book.....L}. In these vacuum gap regions, the electrodynamic force-free criteria break down\,\cite{Goldreich:1969sb} and residual electric fields are non-vanishing. This is also strongly substantiated by pulsar observations, where the vacuum gap regions, and prevalent electric fields there, are thought to play a prominent role in driving pulsar radio emissions\,\cite{Michel:1982fj}. For Magnetars, the induced average electric field in the polar vacuum gap regions will be large, given by\,\cite{2004hpa..book.....L}
\begin{equation}
E\simeq \frac{1}{2}\Omega B R \sim 10^{14} \,\mathrm{V/m}  \; .
\label{eq:EfieldNS}
\end{equation}
Here, $\Omega$, $B$ and $R$ are the rotational velocity, magnetic field and radius of the neutron star. This electric field in the polar vacuum gap region is mostly parallel to the magnetic field there\,\cite{Becker:2009, 2004hpa..book.....L}.

In the presence of such large electric fields, light particles, such as neutrinos with a tiny electric charge, may be non-perturbatively pair produced; via the Schwinger mechanism. For homogeneous fields, where $\vec{B} \shortparallel \vec{E}$, the neutrino pair-production rate per unit volume\,\cite{Nikishov:1970br, Kim:2003qp} is given by
\begin{equation}
\Gamma_{\nu\bar{\nu}}= \frac{Q_\nu^2 E B}{4\pi^2 } \coth\left[\frac{ \pi  B}{E}\right] \exp\left[-\frac{ \pi m_\nu^2 }{ \, Q_\nu E}\right] \; .
\label{eq:eb_spp}
\end{equation}
This may be derived readily\,\cite{Korwar:2018euc} using worldline instanton techniques\,\cite{Affleck:1981ag,DunneWorldline}. Note that for the electric field values and viable neutrino charges of possible interest, the rate is highly suppressed when $m_\nu \gtrsim 1\,\mathrm{eV}$,
but we are outside this regime,
because neutrino masses are constrained below this value.
The most robust constraint is the model-independent bound obtained recently through the
measurement of the end-point of the electron spectrum of
tritium $\beta$-decay in the KATRIN experiment~\cite{Aker:2019uuj}:
$m_\nu < 1.1 \, \mathrm{eV} $ at 90\% CL.
Stronger constraints of the order of $m_\nu \lesssim 0.1\,\mathrm{eV}$
have been obtained from cosmological measurements
in the standard $\Lambda$CDM cosmological model.
Let us also note that
the mass-squared measurements in neutrino oscillation experiments
imply that, except possibly for one mass-eigenstate that may be massless or very light, $m_\nu \gtrsim 0.01\,\mathrm{eV}$. Thus, we may expect that the corresponding neutrino Compton wavelengths are plausibly such that the field homogeneity assumption in Eq. (\ref{eq:eb_spp}) is satisfied to a good degree in the vacuum gap region.

The main energy reservoir of a Magnetar is the super-strong electromagnetic field\,\cite{1992ApJ...392L...9D,1993ApJ...408..194T,Thompson:1995gw}, which is thought to drive the persistent luminosities and burst activities. The pair-production of charged neutrinos will sap energy from this reservoir. This causes a gradual depletion in the electromagnetic energy stored. Based on the observational evidence for Magnetars, with currently deduced characteristic life-times, we may hence leverage a broad energy-balance argument to put conservative limits on $Q_\nu$. This limit will apply to all neutrino flavours and should be model independent, depending only on the fact that a particle with an electric charge will couple to the $U(1)_{\text{\tiny{QED}}}$ gauge field. Similar considerations have already placed interesting limits on generic milli electrically and magnetically charged particles that may exist in nature\,\cite{Hook:2017vyc,Korwar:2017dio}.

Comparing the average power expended by the Magnetar over its active lifetime, to the average power expended for non-perturbative neutrino production and powering the persistent luminosity, one gets the approximate inequality
\begin{equation}
\int dV \left(\frac{d^2\mathbb{E}_{\text{\tiny{lum.}}}}{dt\,dV}+\frac{d^2\mathbb{E}_{\nu \bar{\nu}}}{dt\,dV} \right) \lesssim  \langle P_{\text{\tiny{M}}} \rangle \; ,
\label{eq:eloss} 
\end{equation}
where we have defined
\begin{equation}
 \langle P_{\text{\tiny{M}}} \rangle \equiv  \Big \langle \frac{1}{T} \int dV\,\frac{1}{2}\left( B^{2}+E^{2}\right) \Big \rangle_{\text{\tiny{Magnetar}}} \; .
\end{equation}
Here, $\mathbb{E}_{\text{\tiny{rad.}}}$ and $\mathbb{E}_{\nu \bar{\nu}}$ are the average energies expended in maintaining the luminosities and Schwinger pair production of the lightest $\nu \bar{\nu}$ pairs. $T$ is the average active life-time of a typical Magnetar. 

Now, one has for the pair-production contribution
\begin{equation}
\frac{d^2\mathbb{E}_{\nu\bar{\nu}}}{dt~dV}= \Gamma_{\nu \bar{\nu}} \,Q_\nu E l \; .
\label{eq:spp_work}
\end{equation}
This includes the average power expended for actual pair-production, as well as for the field to accelerate one of the neutrinos out to a characteristic distance $l$.

 A rough estimate for the average radiation component may be taken based on the persistent quiescent X-ray emissions\,\cite{Olausen:2013bpa}
\begin{equation}
 \langle P_{\text{\tiny{lum.}}} \rangle \equiv\Big\langle\int dV\, \frac{d^2\mathbb{E}_{\text{\tiny{lum.}}}}{dt dV}\Big\rangle~\sim~ 10^{27}\,\mathrm{J/s} \; .
\label{eq:pqxrest}
\end{equation}
This will clearly be an underestimate of the total luminosity, and additions to this estimate will only further strengthen the bound. Typically, $ \langle P_{\text{\tiny{lum.}}} \rangle <  \langle P_{\text{\tiny{M}}} \rangle$.

 From Eqs.(\ref{eq:EfieldNS})-(\ref{eq:pqxrest}), we have
 \begin{equation}
 \int dV \frac{Q_\nu^3 E^2 B l}{4\pi^2 } \coth\left[\frac{ \pi  B}{E}\right] \exp\left[-\frac{ \pi m_\nu^2 }{ \, Q_\nu E}\right] \lesssim   \langle P_{\text{\tiny{M}}} \rangle - \langle P_{\text{\tiny{lum.}}} \rangle \; .
 \label{eq:energyineqform}
\end{equation}

To make estimates, we may try typical ranges for the Magnetar parameters\,\cite{Mereghetti:2008je,Olausen:2013bpa,Ozel:2016oaf}---$B \in [10^{14},\,10^{16}]\,\mathrm{G}$, $R\in[10,\,15]\,\mathrm{Km}$, $\tau_{\text{\tiny{period}}} \in [3,\,10]\,\mathrm{s}$, $l\in[10,\,50]\,\mathrm{Km}$ and $T\in[10^3,\,10^5]\,\mathrm{yrs}$. The polar vacuum gap region, over which pair production of charged neutrinos will be active, may also be roughly estimated\,\cite{Korwar:2018euc}, and it comes to be about $\mathcal{O}(0.1)\, \mathrm{Km}^3$. Using these values, numerically solving Eq.\,({\ref{eq:energyineqform}}), one obtains a flavor-universal bound in the range
\begin{equation}
|Q_\nu| ~ \lesssim  ~10^{-11}-10^{-12} e  \; .
\label{eq:elossbound}
\end{equation}

There are astrophysical unknowns and uncertainties present in the Magnetar parameters. The variations due to these on the $Q_\nu$ bound are nevertheless somewhat diluted, in many parametric regions, as we see above. This is because, roughly speaking, in regions where the exponent in Eq. (\ref{eq:eb_spp}) is $\mathcal{O}(1)$, the left hand side of Eq.\,(\ref{eq:energyineqform}) is $\propto Q_\nu^3$ and the consequent variation in the bound on $Q_\nu$ gets reduced by a cube-root. The crude bound in Eq.\,(\ref{eq:elossbound}) is relatively robust in this sense.

This crude bound is solely motivated by the observational evidence for Magnetars, with large magnetic fields and the observed characteristic life-times. Note that this limit is comparable to the bounds deduced in \cite{Gninenko:2006fi,Studenikin:2013my,Parada:2019gvy}, from neutrino-electron scattering and experiments attempting to measure a neutrino magnetic moment. It nevertheless is weaker than the interesting limit in\,\cite{Studenikin:2012vi} represented in Eq.\,(\ref{eq:nust}), based on the  $\nu$ST mechanism, and the stellar cooling bounds of Eqs. (\ref{eq:shb}) and (\ref{eq:scb}). The former bound from neutron stars was obtained by modeling the propagation of neutrinos out through the magnetized rotating nuclear media, and solving the relevant Dirac equation with that ansatz.

\section{Summary and conclusions}
\label{sec:summary}

The question of a possibly small neutrino charge is an intriguing one.
In the present work we revisited some aspects of this problem and pointed out two new direct
bounds from scattering, as well as a lower and an upper bound
that are motivated, respectively,
by the weak gravity conjecture and by the observation of Magnetars. 

We reviewed the little-known proof that
the quantum field theoretic consistency of the full three-generation Standard Model
with massless neutrinos allows them to have electric charges,
albeit with the restriction
that two neutrinos can have opposite charges,
while the third must be neutral.
This is related to the presence of a non-anomalous abelian symmetry
generated by one of the three differences of the generation lepton numbers.
The addition of right-handed neutrinos and the corresponding Dirac mass terms in the Lagrangian, giving neutrinos Dirac masses, also lead to charge dequantization and non-zero neutrino charges. 

There exists strong experimental and observational constraints on the neutrino charges. We reviewed some of these constraints, pointing out some relations, as well as making additions to the set. Specifically,
we obtained new direct upper bounds on the electric charges
of $\nu_\mu$ and $\nu_\tau$
from the most stringent experimental bounds on the corresponding magnetic moments.
We also utilized the weak gravity conjecture and observational evidence for Magnetars to add to the set of strong constraints already existing in literature. 

For the latter set of constraints, firstly by leveraging the hypothesis that gravity is the weakest force (the essence of the weak gravity conjecture), we were able to motivate a possible lower-bound on a non-zero neutrino charge, if it exists. Secondly, the observation of highly magnetic neutron stars (Magnetars) suggested that charged neutrinos must be non-perturbatively pair produced in certain regions of the stellar atmosphere. This would contribute to the depletion of the electromagnetic energy reservoir of the neutron star, and hence by energetic arguments also place novel upper bounds on the neutrino charge. The limit thus obtained is found to be comparable to recent limits derived from reactor neutrinos, through the analysis of elastic electron-neutrino cross sections\,\cite{Parada:2019gvy}. Limits on neutrino charges derived from experiments constraining neutrino magnetic moments are also in a similar ballpark\, \cite{Gninenko:2006fi,Studenikin:2013my}.
 

\section*{Acknowledgments}

The work of A. Das is supported by the JSPS, Grant-in-Aid for Scientific Research, No. 18F18321. D. Ghosh acknowledges support through the Ramanujan Fellowship and the MATRICS grant of the Department of Science and Technology, Government of India. The work of C. Giunti was partially supported by the research grant ``The Dark Universe: A Synergic Multimessenger Approach" number 2017X7X85K under the program PRIN 2017 funded by the Ministero dell'Istruzione, Universit\`a e della Ricerca (MIUR). A. Thalapillil would like to acknowledge support from an Early Career Research award from the Department of Science and Technology, Government of India. C. Giunti and A.Thalapillil would also like to thank the organizers of the Gordon Research Conference on Particle Physics 2019, where this study was initiated.


\bibliography{NeutrinoChargeConstraints}

\end{document}